\documentclass[10pt]{IEEEtran}

\usepackage{graphicx}
\usepackage{color}

\usepackage{cite}

\usepackage{amsbsy}  
\usepackage{em2}     
\usepackage{specfun}

\newcommand{\eqref}[1]{(\ref{#1})}
\newcommand{\eqspairref}[2]{(\ref{#1}),(\ref{#2})}

\newcommand{\Eqref}[1]{Equation~(\ref{#1})}

\newcommand{\figref}[1]{figure~\ref{#1}}
\newcommand{\Figref}[1]{Figure~\ref{#1}}

\newcommand{\Hantila}{H\v{a}n\c{t}il\v{a} }

\newcommand{\setzerospace}{\setlength{\arraycolsep}{0.0em}}

\setzerospace

\newcommand{\figwidth}{8cm}

\begin{document}


\title{A modal approach for the solution of the non-linear induction problem in ferromagnetic 
media}
\author{Anastassios~Skarlatos and~Theodoros~Theodoulidis,~\IEEEmembership{Senior Member,~IEEE}
\thanks{A. Skarlatos is with the CEA, LIST, F-91191 Gif-sur-Yvette cedex, France.}
\thanks{T. Theodoulidis is with the Department of Mechanical Engineering, University of 
Western Macedonia, Bakola \& Sialvera, 50100 Kozani, Greece.}
}

\maketitle


\begin{abstract}
The non-linear induction problem in ferromagnetic media is solved using the fixed-point 
iteration method, where the linearized problem at each iteration is treated by means of a 
modal approach. The proposed approach does not require meshing of the solution domain, which 
results in fast computations comparing to conventional mesh-based numerical techniques. Both 
harmonic and pulse excitations are considered via Fourier and Laplace transform, respectively. 
An efficient method for the fast computation of the inverse Laplace transform of the magnetic 
polarization signals is also devised based on the generalized pencil-of-function (GPOF) 
method. Although being restricted to one dimensional configurations, the present work provide 
the tools for the treatment of two and three dimensional problems, whose study is under way.
\end{abstract}

\section{Introduction}

The solution of the induction problem in configurations involving ferromagnetic media, is in 
general case a non-linear problem. For relatively weak fields, linearization provides a very 
good approximation of the real solution. This is for example the case in classical 
eddy-current testing (ECT) or wave propagation applications. Nonetheless, when excitation 
currents of medium or high intensity are involved, the full non-linear formulation has to be 
considered. Such currents are encountered in electric machines and transformers as well as 
in material evaluation applications, just to mention some of the most pertinent practical 
situations.

The general way of addressing a non-linear field problem can be roughly sketched  as follows: 
construct a sequence of successive approximations to the solution by treating a linearized 
problem each time, and iterate until a convergence criterion is satisfied. There are 
principally two main approaches for establishing the above iterative scheme; either by 
expressing the non-linear formulation in terms of a fixed-point problem, that is by 
constructing the sequence $\vect{x}^{(l+1)} = \mathcal{G}\vect{x}^{(l)}$, or by reformulating 
it as an homogeneous equation $\mathcal{F}\vect{x}=0$ and searching its roots using the 
Newton-Raphson method, i.e. via the relation $\vect{x}^{(l+1)} = \vect{x}^{(l)}-
\left[d\mathcal{F}/d\vect{x}^{(l)}\right]^{-1}\mathcal{F}\vect{x}^{(l)}$. Both methods have 
pros and cons regarding their stability and their convergence speed, and various techniques 
based on the introduction of relaxation conditions as well as hybridization of the two 
approaches have been proposed for achieving optimal performance \cite{munteanu_compel01}.

Considerable effort has been given during the past years in applying both iterative schemes
for treating the electromagnetic induction problem in ferromagnetic media, where the 
linearized problem at each iteration is solved by means of a numerical technique. In most 
cases, the finite elements method (FEM) is the numerical tool of choice 
\cite{albanese_transmag92,biro_transmag06}, yet numerous works using other mesh-based 
approaches like the finite integration technique (FIT) \cite{drobny_transmag00} or the 
integral equation approach \cite{albanese_transmag96,circ_transmag07} exist as well. The 
relative literature is vast, and the above list of references tends only to be indicative.  

The application of analytical methods for the solution of the non-linear field problem is by 
far less studied, to the best of the authors knowledge. This trend may be explained by the 
difficulty of the problem, and the limited number of cases that can be tackled using 
analytical approaches in comparison with the existing versatile numerical tools. Yet, a few 
attempts to address simple geometries, usually by resorting to (sometimes coarse) 
approximations, can be found in the literature. Such attempts comprise the approximation of
the abrupt magnetic transition approximation, initially proposed by Wolman and Kaden 
\cite{wolman_ztechnphys35} and further worked in later articles \cite{maclean_japphys54,
papoulis_japphys54}, or the treatment of the eddy-current problem in a ferromagnetic rod 
excited by a long solenoid by means of a perturbation approach as is proposed by Nayfeh and 
Asfar in \cite{nayfeh_transmag74}. The cartesian counterpart of the latter problem, namely an 
infinite plate with uniform current layers on both sides is treated by Labridis and Dokopoulos 
in \cite{labridis_transmag89} by performing successive approximations to the magnetic 
permeability in the diffusion equation.

An alternative strategy to such ad hoc approaches with limited domain of validity, would be to 
apply a modal-based approach for the inversion of the linearized problem occurring at each 
iteration of the general iterative techniques mentioned above. Modal techniques have proved to 
be a very powerful tool, which allowed the treatment of a number of important linear 
eddy-current problems \cite{burke_jphysicsD04,bowler_jphysd05,theodoulidis_procrsoca05,
theodoulidis_jap08,skarlatos_procrsoca12}. The interest in applying modal methods, relies on 
the shorter calculation times with respect to the numerical techniques, the absence of the 
computational grid as well as the better insight to the physics of the problem they provide.

The inherit difficulty in treating inhomogeneous materials using modal techniques, makes their 
use with Newton-Raphson like schemes, where the Jacobian of the material law is involved, 
complicated. This restriction, however, does not occur when considering fixed-point schemata 
which leave the material properties of the linerized problem intact. This in conjunction with 
its good stability properties makes the fixed point approach the method of choice in this work.

The main purpose of this article is to explore the construction of such semi-analytical based 
solutions to the non-linear induction problem. Since it is a first attempt, the focus in this 
article will lie on setting up a solving strategy with regard to two rather academic problems, 
namely a one-dimensional coil with a ferromagnetic core in the cartesian and the cylindrical 
systems. Yet, despite being of very simple geometry, these problems still present a practical 
interest, since the herein developed solutions may deliver some useful approximations for 
configurations met in engineering problems like yokes and coil cores.

Both harmonic and pulsed excitations will be considered in the context of this work. The 
discrete spectrum of the former makes the frequency domain formulation in combination with 
Fourier synthesis the most straight-forward and, in the same time, efficient approach 
\cite{biro_transmag06,circ_transmag07}. In the latter case, the direct treatment in time 
domain seems to be more relevant. Following previous works, the analysis is facilitated by 
formulating the problem in the Laplace domain \cite{bowler_transmag97,fu_transmag06}. A 
considerable difficulty when working in the Laplace domain is the inverse transformation of 
arbitrary signals back into the time domain. Instead of adopting a numerical inversion, an 
alternative approach is proposed based on the generalized pencil-of-function (GPOF) method 
\cite{jain_tansjassp83,hua_tansap89}. According to the GPOF method, an arbitrary signal can be 
approximated in the Laplace domain by a sum of rational functions, whose inverse 
transformation back into the time domain exist in closed form. 

Work is under way for the extension of the herein developed ideas to two dimensional 
configurations, in order to tackle more realistic configurations involving cylindrical coils 
around bars or above ferromagnetic plates. 

\section{Formal solution using the fixed-point method}

The magnetic constitutive relation of a ferromagnetic material reads:
%
\begin{equation}
\Bflux = \permb_0[\Hfield + \Magn(\Hfield)]
\label{eq:ConsRelVec}
\end{equation}	
where $\permb_0$ is the magnetic permeability of the free space, and $\Magn$ stands for the 
magnetization of the material, which comprises the non-linear effects. An equivalent, in 
some cases more useful form of \eqref{eq:ConsRelVec}, is obtained by introducing the magnetic 
polarization $\Ipol:=\permb_0\Magn$. \Eqref{eq:ConsRelVec} becomes then
%
\begin{equation}
\Bflux = \permb_0\Hfield + \Ipol(\Hfield).
\label{eq:ConsRelPol_B}
\end{equation}
The last equation can be written formally as follows
%
\begin{equation}
\Ipol\fun{\Bflux} = \Bflux - \permb_0\Hfield\fun{\Bflux}
\label{eq:ConsRelPol_I}
\end{equation}
where $\Hfield\fun{\Bflux}$ is the inverse material curve in the $\Hfield-\Bflux$ plane, which
for the purposes of our analysis will be considered known. 

For reasons that will be explained below, it is interesting to consider a magnetic 
permeability different than that of the free space, i.e. $\permb>\permb_0$. The magnetic 
polarization in \eqref{eq:ConsRelPol_B} and \eqref{eq:ConsRelPol_I} then should be replaced by 
an effective one, namely $\Ipol\rightarrow\Ipol'+ \left(\permb-\permb_0\right)\Hfield$. Notice 
that, $\Ipol'$ in contrast with the original variable $\Ipol$, does not represent a physical 
quantity any more; it is merely introduced for mathematical convenience. Once $\Ipol'$ defined, 
the prime symbol will be dropped henceforth, i.e. the use of the effective magnetic 
polarization will be implied from this point forward.

Using \eqref{eq:ConsRelPol_I} and applying the quasi-static approximation, the diffusion 
equation for the magnetic field in the magnetic medium becomes
%
\begin{equation} 
\nabla\times\nabla\times\Bflux + \permb\cond\frac{\partial \Bflux}{\partial t} = 
\nabla\times\nabla\times\Ipol\fun{\Bflux}.
\label{eq:CurlCurl}
\end{equation}
Taking into account $\nabla\cdot\Bflux=0$, and making the plausible assumption of the absence 
of free magnetic poles in the bulk material ($\nabla\cdot\Ipol=0$)\footnote{Notice that this 
assumption is justified only in the case of homogeneous $\permb$.}, \eqref{eq:CurlCurl} 
reduces to 
%
\begin{equation}
\nabla^2\Bflux - \permb\cond\frac{\partial \Bflux}{\partial t} = \nabla^2\Ipol\fun{\Bflux}.
\label{eq:InhomHelmhVect}
\end{equation}
Let us set $\mathcal{D}:=\nabla^2 - \permb\cond\partial/\partial_t$, the diffusion operator. 
According to Banach fixed point theorem, if the non-linear operator 
$\mathcal{D}^{-1}\nabla^2\Ipol$ is a contraction, then the sequence
%
\begin{equation}
\Bflux^{(l+1)} = \mathcal{D}^{-1}\nabla^2\Ipol\dfun{\Bflux^{(l)}}
\label{eq:FixedPointIter}
\end{equation}
converges to the fixed-point of the operator, which is also the solution of 
\eqref{eq:InhomHelmhVect}. The condition for the convergence of \eqref{eq:FixedPointIter} is 
that the magnetization curve is Lipschitzian and monotonous, as it has been shown by \Hantila 
in \cite{hantila_revroum75}. This condition is met by the sigmoid B-H curves of the most 
important ferromagnetic media for $\permb<2\permb_{min}$, $\permb_{min}$ being the minimum of 
the differential permeability. The closer the value of $\permb$ to the upper limit, the better 
the contraction factor, which justifies our interest to work with the effective magnetic 
polarization introduced above. Practically, the choice of $\permb$ is a compromise between 
convergence speed and stability of the algorithm. For the rest of the article, the convergence 
will be considered as guaranteed without further investigation.

The general iterative algorithm for the solution of the non-linear equation using the 
fixed-point theorem, an approach referred also to in the literature as Picard-Banach or 
polarization method, is summarized schematically by the following pseudo-code

\begin{center}
\begin{tabular}{l}
\verb+ Set+ $\Ipol^{(0)},\Delta\Ipol^{(0)}$ \\
\verb+ for+ $l=0,1,\ldots$ \verb+ until+ 
$\left|\left|\Delta\Ipol^{(l)}\right|\right|<\epsilon$ \\
\hspace{4ex}\verb+ Solve+ $\Bflux^{(l+1)} = \mathcal{D}^{-1}\nabla^2\Ipol^{(l)}$ \\
\hspace{4ex}\verb+ Evaluate+ 
$\Ipol^{(l+1)} = \Bflux^{(l+1)} - \permb\Hfield\dfun{\Bflux^{(l+1)}}$ \\
\hspace{4ex}\verb+ Calculate the residual+ $\left|\left|\Delta\Ipol^{(l+1)}\right|\right|$
\end{tabular}
\end{center}

The above scheme is generic in the sense that it does not depend on the way that 
$\mathcal{D}^{-1}$ is calculated. In the following paragraphs the modal approach will be 
applied for the solution of \eqref{eq:InhomHelmhVect} for two important piece geometries: 
infinite plate and cylindrical rod.

\section{1D cartesian problem: ferromagnetic plate between two infinitely long current layers}

\begin{figure}[h]
\centering
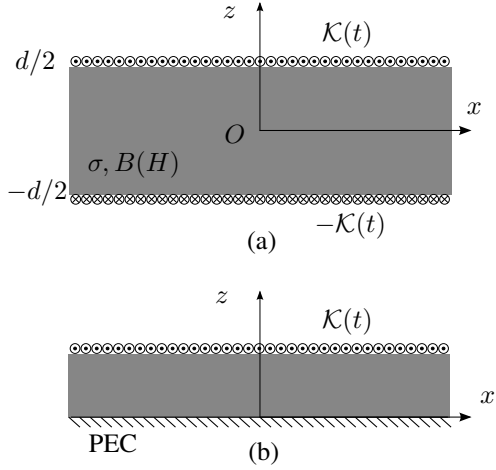
\caption{Infinitely long ferromagnetic plate between two opposite current layers: (a) real 
configuration and (b) equivalent structure with odd parity.}
\label{fig:cart:conf}
\end{figure}

We shall begin the study of the non-linear eddy-current problem by considering a ferromagnetic 
plate of thickness $d$ sandwiched by two opposite infinite current layers $\mathcal{K}\fun{t}$. 
The geometry of the problem is illustrated in \figref{fig:cart:conf}. Despite its simplicity,
this 1D configuration resembles an infinite, transnationally symmetric yoke, and hence the 
derived solution can be used to approximate the behaviour of realistic 3D yokes.    
%

As it is shown in \figref{fig:cart:conf}, the antisymmetry of the configuration allows the 
problem to be reformulated in an equivalent one, where the solution is restricted to the 
non-negative $z$ semi-axis in combination with a perfectly electric conductor boundary 
condition at the $z=0$ plane (plane of symmetry). This reformulation results in a 
simplification of the original problem.

Owing to the symmetry of the configuration, the magnetic field will lie along the $x$ 
direction and it will depend only upon the distance from the interface, i.e. it will be 
$\Hfield = H\fun{z}\uvect{x}$, whereas the electric field will be directed along $y$ and it 
will be also a function of $z$, namely $\Efield = E\fun{z}\uvect{y}$. Consequently, the vector 
equation \eqref{eq:InhomHelmhVect} reduces to the scalar one
%
\begin{equation}
\frac{\partial^2 B}{\partial z^2} - \permb\cond\frac{\partial B}{\partial t} = 
\frac{\partial^2 I}{\partial z^2}.
\label{eq:cart:InhomWave}
\end{equation}
The magnetic constitutive relation that we need to consider is also scalar, i.e. we can 
write
%
\begin{equation}
I\fun{B} = B - \permb H\fun{B}.
\label{eq:cart:ConsRel}
\end{equation}

In the following two subsections we shall derive the solution to the one dimensional cartesian 
problem for the two most interesting types of excitation, namely harmonic and unit step. Other 
more general forms of excitations like the periodic waveforms (square, triangular etc.) or 
pulse (with rectangular or exponential profile) can be derived based on the solutions of the 
basic aforementioned cases. For the analysis of periodic signals the use of the Fourier series 
is better suited, whereas the Laplace transform is the most efficient tool when studying 
plused excitations. Hence, the analysis will be adapted to the type of the problem that we are 
studying in order to use the most efficient approach.  

\subsection{Harmonic excitation}

Let us assume a harmonic current excitation, i.e. let $\mathcal{K}\fun{t} = 
\mathcal{K}_0\cos\fun{\omega_0 t}$ with $\omega_0 = 2\pi f_0$ being the angular frequency. 
Since we are dealing with a non-linear problem the final signals will still be periodic, yet 
their spectrum will consist by an infinite number of frequencies that are integer multiples of 
the excitation one $f_0$, i.e. it will be $f_n = n f_0$ with $n=1,2,\ldots\infty$\footnote{
It is a well known feature of magnetic materials that, owing to the symmetry of the B-H curve, 
only odd harmonics deliver a non-zero contribution to the solution. Nonetheless, numerical 
experimentation has revealed that taking also the even harmonics into account leads to 
slightly better results. This could be possibly explained by the fact that the actual B 
solution calculated at every iteration is not perfectly symmetric.}. The magnetic induction 
can be described by means of a Fourier series, namely
%
\begin{equation}
B\fun{z,t} = \sum\limits_{n=-\infty}^{n=\infty} B_n\fun{z} e^{\iota n\omega_0 t}
\label{eq:cart:Fourier}
\end{equation}
with the series coefficients being given by
%
\begin{equation}
B_n\fun{z} = \frac{1}{T}\int\limits_{0}^{T} B\fun{z,t} e^{-\iota n\omega_0 t} dt
\label{eq:cart:FourierCoeff}
\end{equation}
and $T=1/f_0$ being the period of the signal. The corresponding expansions for $H$ and $I$ are 
given by the same relations.

The diffusion equation \eqref{eq:cart:InhomWave} reduces then to the Helmholtz inhomogeneous 
equation, which for $n$-th harmonic reads
%
\begin{equation}
\frac{d^2 B_n}{d z^2} - k_n^2 B_n = \frac{\partial^2 I_n}{\partial z^2}
\label{eq:cart:Helmholtz}
\end{equation}
with $k_n^2 = \iota n\omega_0\permb\cond = \iota\omega_n\permb\cond$. 

Assume now that the magnetic polarization $I_n$ is known. Then following the standard theory 
of differential equations, the solution of \eqref{eq:cart:Helmholtz}, can be constructed by 
the superposition of the solution to the homogeneous equation and a partial solution. The 
former, after taking symmetry of the tangential magnetic field across the $z=0$ plane into 
account proves easily to be
%
\begin{equation}
B_n^{(h)}\fun{z} = A_n \cosh\fun{k_n z}.
\end{equation}
The expansion coefficient $A_n$ will be determined by the application of the boundary 
conditions at $z=d/2$. 

In order to construct the partial solution of \eqref{eq:cart:InhomWave}, we assume that at a 
given time instant $t$, both the magnetic field and the polarization can be expanded in the 
eigenfunction-basis of the spatial operator $\partial^2/\partial_z^2$, i.e. 
%
\setzerospace
\begin{equation}
\left\{
\begin{array}{c}
I_n\fun{z} \\
B^{(p)}_n\fun{z}
\end{array}
\right\}
=
\left\{
\begin{array}{c}
C_{n,0} \\
D_{n,0}
\end{array}
\right\}
+
\sum\limits_{i=1}^{\infty} 
\left\{
\begin{array}{c}
C_{n,i} \\
D_{n,i}
\end{array}
\right\}
\cos\fun{\kappa_i z}
\label{eq:cart:HMexp}
\end{equation}
with the eigenvalues
%
\begin{equation}
\kappa_i = (2i-1)\frac{\pi}{d}, \; i=1,\ldots\infty.
\end{equation}
The $C_{n,i}$ development coefficients are obtained by the projection of the magnetic
polarization on the modes, namely
%
\begin{equation}
C_{n,i} =
\left\{
\begin{array}{ll}
\frac{1}{d}
\int\limits_{0}^{d/2} 
I_n\fun{z} dz, 
&$ for $ i=0\\
\frac{2}{d}
\int\limits_{0}^{d/2} 
I_n\fun{z}
\cos\fun{\kappa_i z}dz,
&$ for $ i=1,2,\ldots\\
\end{array} 
\right.
\end{equation}
The expansion coefficients for the magnetic induction $D_{n,i}$ are determined by substituting 
\eqref{eq:cart:HMexp} into \eqref{eq:cart:InhomWave}, which yields
%
\begin{equation}
D_{n,i} =
\left\{
\begin{array}{ll}
0, 
&$ for $ i=0\\
\frac{\kappa_i^2}{\kappa_i^2 + k_n^2} C_{n,i},
&$ for $ i=1,2,\ldots\\
\end{array} 
\right.
\label{eq:cart:DniCoef}
\end{equation}

The general solution of \eqref{eq:cart:InhomWave} for the $n$-th harmonic is thus given by
\begin{equation}
B_n\fun{z} = A_n \cosh\fun{k_n z} + 
\sum_{i=1}^{\infty} \frac{\kappa_n^2}{\kappa_i^2 + k_n^2} C_{n,i}\cos\fun{\kappa_i z}.
\label{eq:cart:ihnom:FormSol}
\end{equation}
Applying the boundary condition for the magnetic field at the upper surface of the plate to 
\eqref{eq:cart:ihnom:FormSol} and taking \eqref{eq:cart:ConsRel} into account, we obtain
%
\begin{equation}
\frac{1}{2}\mathcal{K}_0\delta_{n,\pm 1} = 
\permb^{-1} 
\left[
A_n \cosh\fun{\frac{k_n d}{2}} + C_{n,0}
\right]
\label{eq:cart:Acoef}
\end{equation}
where $\delta$ is the Kronecker's delta. 

Therefore, the solution for the $n$-th harmonic of the magnetic field becomes at $(l+1)$th
iteration
%
\begin{eqnarray}
B_n^{(l+1)}\fun{z} &= 
\left[
\frac{\permb}{2}\mathcal{K}_0\delta_{n,\pm 1} - C_{n,0}^{(l)}
\right]
\frac{\cosh\fun{k_n z}}{\cosh\fun{k_n d/2}}
\nonumber\\
&+
\sum\limits_{i=1}^{\infty} \frac{\kappa_n^2}{\kappa_i^2 + k_n^2} C_{n,i}^{(l)}\cos\fun{\kappa_i z}.
\end{eqnarray}
The first term of the sum gives the response of the linearized medium to the current, whereas
the remaining two terms yield the contribution of the magnetic polarization. Note that the
development coefficients for the magnetic polarization are calculated based on its value at
the $l$th iteration, following the Picard-Banach scheme. The final expression in time domain
will be obtained by summing the contributions of all harmonics according to
\eqref{eq:cart:Fourier}.

\subsection{Pulsed excitation}

Let us now consider a current described by an arbitrary pulse $\mathcal{K}\fun{t}$. Our main 
tool here will be the Laplace transform $F\fun{s} = \mathcal{L}\left[f\fun{t}\right]$, defined
by the integral pair
%
\begin{equation}
F\fun{s} :=
\int\limits_0^\infty f\fun{t} e^{-st} dt
\label{eq:LaplaceTransf}
\end{equation}
for the forward transform, and
%
\begin{equation}
f\fun{t} =
\frac{1}{2\pi \iota}\int\limits_{Br} F\fun{s} e^{st} ds
\label{eq:InvLaplaceTransf}
\end{equation}
for the inverse transform, where $Br$ stands for the Bromwich path of integration in the 
complex $s$-plane.

Taking the Laplace transform of \eqref{eq:cart:InhomWave} we obtain the inhomogeneous 
Helmholtz equation in the $s$-plane
%
\begin{equation}
\frac{\partial^2 B}{\partial z^2} - s\permb\cond B = \frac{\partial^2 I}{\partial z^2}.
\label{eq:cart:LHelmholtz}
\end{equation}
The solution of the homogeneous equation with account for the symmetry at $z=0$ reads
%
\begin{equation}
B^{(h)}\fun{z,s} = A\fun{s} \cosh\fun{\sqrt{\permb\cond s} z}
\end{equation}
the coefficient $A\fun{s}$ being determined by the excitation waveform and the boundary 
conditions. 

For the construction of the partial solution, we work as in the harmonic case, i.e. the 
spatial part of the magnetic field and the magnetic polarization is expanded in the 
eigenfunction basis \eqref{eq:cart:HMexp} yielding the following expressions for the 
development coefficients:
%
\begin{equation}
C_i\fun{s} =
\left\{
\begin{array}{ll}
\frac{1}{d}
\int\limits_{0}^{d/2} 
I_n\fun{z,s} dz, 
&$ for $ i=0\\
\frac{2}{d}
\int\limits_{0}^{d/2} 
I_n\fun{z,s}
\cos\fun{\kappa_i z}dz,
&$ for $ i=1,2,\ldots\\
\end{array} 
\right.
\label{eq:CisCoeff}
\end{equation}
and
%
\begin{equation}
D_i(s) = \frac{1}{1 + s\tau_i} C_i\fun{s}
\label{eq:cart:CoeffSol}
\end{equation}
with $\tau_i = \permb\cond/\kappa_i ^2$.

Application of the tangential magnetic field continuity across the plate surfaces yields 
%
\begin{equation}
A\fun{s} \cosh\fun{\sqrt{\permb\cond s} d/2} = 
\permb\mathcal{K}\fun{s} + C_0\fun{s}
\end{equation}
which implies for the total solution of \eqref{eq:cart:LHelmholtz}
%
\begin{eqnarray}
B\fun{z,s} &= 
\left[\permb\mathcal{K}\fun{s} + C_0\fun{s}\right]
\frac{\cosh\fun{\sqrt{sT}\;z/d}}{\cosh\fun{\sqrt{sT}/2}} 
\nonumber\\
&+
\sum\limits_{i=1}^{\infty}\frac{1}{1 + s\tau_i} C_i\fun{s}\cos\fun{\kappa_i z}
\label{eq:cart:TotalSol}
\end{eqnarray}
where we have set $T:=\permb\cond d^2$. Note that $T$ is a measure of the system characteristic
time \cite{jacksonbook}.

The functional form of $C_i\fun{s}$ is arbitrary. Using the generalized pencil-of-function 
(GPOF) method \cite{jain_tansjassp83,hua_tansap89} though, we can approximate it via a 
weighted sum of rational functions, deduced by its dominant poles in the complex plane, namely
%
\begin{equation}
C_i\fun{s}\approx \sum\limits_{j=1}^{N_p} \frac{b_{ij}}{s-p_{ij}}
\end{equation}
and \eqref{eq:cart:TotalSol} reduces to the more convenient form
%
\begin{eqnarray}
B\fun{z,s} &= \left[
\permb\mathcal{K}\fun{s} + 
\sum\limits_{j=1}^{N_p} \frac{b_{0j} }{s-p_{0j}} \right]
\frac{\cosh\fun{\sqrt{sT}\;z/d}}{\cosh\fun{\sqrt{sT}/2}}
\nonumber\\
&+
\sum\limits_{i=1}^{\infty}\frac{1}{1+ s\tau_i}\cos\fun{\kappa_i z}
\sum\limits_{j=1}^{N_p}
\frac{b_{ij}}{s-p_{ij}}.
\label{eq:cart:TotalSolPolesDecomp}
\end{eqnarray}

It is convenient to consider the two contributions, namely that of the excitation current and
the one from the magnetic polarization, separately, i.e. we write
\begin{equation}
B\fun{z,s} = B_c\fun{z,s} + B_p\fun{z,s}.
\end{equation}
We shall work first with the second term since it is independent from the excitation. From
inverse Laplace transform tables, we get the formula \cite[(40), p. 259]{erdelyitabitrans}
%
\begin{equation}
\mathcal{L}^{-1}
\left[
\frac{1}{s-p}\frac{\cosh\fun{x\sqrt{s}}}{\cosh\fun{l \sqrt{s}}}
\right] = 
\frac{\cosh\fun{x\sqrt{p}}}{\cosh\fun{l\sqrt{p}}} e^{pt} 
-
y\fun{\frac{x}{2 l}\left|\frac{t}{l^2}\right| p l^2}
\label{eq:cart:CoshInvLap}
\end{equation}
where
%
\begin{eqnarray}
y\fun{v|\tau|\nu} := 
2\pi\sum\limits_{n=0}^{\infty}
&\frac{(n+1/2)(-1)^n\cos[(n+1/2)2\pi v]}{(n+1/2)^2\pi^2 + \nu}
\nonumber\\
&
\hspace{3ex}
\times e^{-(n+1/2)^2\pi^2 \tau}.
\label{eq:cart:ydef}
\end{eqnarray}

Using \eqspairref{eq:cart:CoshInvLap}{eq:cart:ydef} as well as the standard formulas for the
rational function transforms \cite{erdelyitabitrans}, $B_p$ is transformed into the time domain 
- at the $(l+1)$th iteration - as follows 
%
\begin{eqnarray}
B_p^{(l+1)}\fun{z,t} &= 
\sum\limits_{j=1}^{N_p} b_{0j}^{(l)}
\frac{\cosh\fun{\sqrt{p_{0j} T} \hat{z}}}{\cosh\fun{\sqrt{p_{0j} T}/2}} e^{p_{0j} t}
\nonumber\\
&-
\sum\limits_{j=1}^{N_p} b_{0j}^{(l)}
y\fun{\hat{z}\left| \frac{4 t}{T}\right| p_{0j}T}
\nonumber\\
&+
\sum\limits_{i=1}^{\infty}
\sin\fun{\kappa_i z}
\sum\limits_{j=1}^{N_p}
b_{ij}^{(l)}
\frac{e^{p_{ij} t} - e^{-t/\tau_i}}{1+p_{ij}\tau_i}
\end{eqnarray}
where $\hat{z}:=z/d$ stands for the normalized depth. 

The time domain expression for the current response $B_c$ depends on the excitation waveform, 
and it is not always amenable to a closed form. However, if we consider a step function 
excitation we arrive at a relatively simple expression for $B_c$ using the inverse transform 
\eqref{eq:cart:CoshInvLap}
%
\begin{equation}
B_c\fun{z,t} = 
\permb \mathcal{K}_0 \left[1 - y\fun{\hat{z}\left| \frac{4 t}{T}\right| 0}\right].
\label{eq:cart:BlinTD}
\end{equation}
Perhaps a more elegant expression can be given in terms of the elliptic theta function. 
Recalling the transform \cite[(37), p. 259]{erdelyitabitrans},
%
\begin{equation}
\mathcal{L}^{-1}
\left[
\frac{\cosh\fun{x\sqrt{s}}}{\cosh\fun{l \sqrt{s}}}
\right]
= 
-\frac{1}{l}\frac{\partial}{\partial x}
\thetafun{1}\fun{\frac{x}{2 l}\left|\frac{\iota\pi t}{l^2}\right.}
\label{eq:cart:C0InvLaplaceTheta}
\end{equation}
where $-l\le x \le l$, and $\thetafun{1}$ is the elliptic theta function defined as 
\cite[pp. 387]{erdelyitabitrans}
%
\begin{equation}
\thetafun{1}\fun{v|\tau} = (-\iota\tau)^{-1/2}
\sum\limits_{n=-\infty}^{\infty}
(-1)^n e^{-\iota\pi(v-1/2+n)^2/\tau}
\label{eq:cart:ElThetaFun}
\end{equation}
\eqref{eq:cart:BlinTD} is written as
%
\begin{equation}
B_c\fun{z,t} = 
-\frac{1}{2}\permb\mathcal{K}_0
\int\limits_0^{t/T}
\frac{\partial}{\partial \hat{z}}
\thetafun{1}\fun{\hat{z}\left|\iota\pi\tau\right.}
d\tau.
\end{equation}

Knowing the field response to a step current $B_s$, the respective response to an arbitrary 
excitation $s$ can be easily obtained by utilizing the Duhamel's integral
\begin{equation}
B_c\fun{z,t} = \frac{d}{dt} \int\limits_0^t s\fun{x} B_s\fun{z,t-x} dx.
\label{eq:cart:Duhamel}
\end{equation}

It is interesting to note that the current-related term of the plate response $B_c$ for short 
times reduces to the corresponding solution for a half-space. This is not a surprising result 
since at sufficiently early times, the induced field at each side of the plate has not yet 
reached the opposite side, and the plate behaves as being of infinite thickness. The detailed 
proof is given in the appendix.

\section{1D cylindrical problem: infinite solenoid}

\begin{figure}[h]
\centering
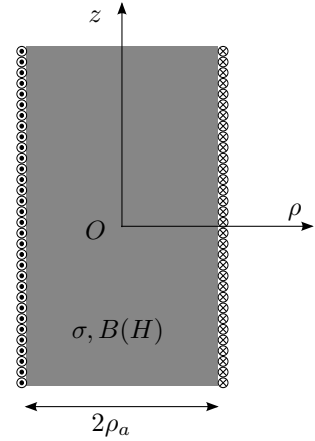
\caption{Infinite solenoid with a ferromagnetic core.}
\label{fig:cyl:conf}
\end{figure}

We move now to the cylindrical geometry by considering an infinite solenoid encircling an 
infinitely long ferromagnetic core. The geometry of the problem is illustrated in 
\figref{fig:cyl:conf}.
%
The radius of the solenoid is $\rho_a$ and the the number of turns per unit length is $N$.  
The core is assumed to be a cylindrical rod of the same dimensions with the solenoid. The 
core material is considered to be again ferromagnetic with the same characteristics as the 
one examined above. The symmetry of the configuration results in a parallel to coil axis 
magnetic field ($\Hfield = H\uvect{z}$) and an azimuthal electric field
($\Efield = E\uvect{\phi}$), and thus the magnetic constitutive relation will be again 
scalar, this time involving $B_z,H_z$ and $\hat{I}_z$\footnote{To avoid confusion 
with the modified Bessel function of the first kind, in the cylindrical problem the magnetic 
polarization will be distinguished by a hat.}. The functional form of the constitutive 
equation will be given by \eqref{eq:cart:ConsRel}.

The wave equation for the magnetic field in cylindrical coordinates reads
%
\begin{equation}
\frac{\partial^2 B}{\partial\rho^2} + \frac{1}{\rho}\frac{\partial B}{\partial\rho}-
\permb\cond \frac{\partial B}{\partial t} = 
\frac{\partial^2 \hat{I}}{\partial\rho^2} + \frac{1}{\rho}\frac{\partial \hat{I}}{\partial\rho}.
\label{eq:cyl:IhnomWave}
\end{equation}

In the next sections we are going to consider separately the two time excitations that we also
treated for the cartesian problem.

\subsection{Harmonic excitation}

The inhomogeneous Helmholtz equation for the $n$-th frequency reads
%
\begin{equation}
\frac{\partial^2 B_n}{\partial\rho^2} + \frac{1}{\rho}\frac{\partial B_n}{\partial\rho}
-k_n^2 B_n = 
\frac{\partial^2 \hat{I}_n}{\partial\rho^2} + \frac{1}{\rho}\frac{\partial \hat{I}_n}{\partial\rho}
\label{eq:cyl:InhomHelmholtz}
\end{equation}
with the wavenumber
%
\begin{equation}
k_n^2 = \iota n\omega_0\permb\cond.
\end{equation}
The solution of the homogeneous equation is given by the zero-order modified Bessel function
%
\begin{equation}
B_n^{(h)}\fun{\rho} = A_n \besseli{0}\fun{k_n \rho}.
\end{equation}

The eigenvalue equation of the spatial operator in \eqref{eq:cyl:InhomHelmholtz} reads
%
\begin{equation}
\left[
\frac{\partial^2}{\partial\rho^2} + \frac{1}{\rho}\frac{\partial}{\partial\rho}
\right]
\besselj{0}\fun{\kappa\rho} = 
-\kappa^2\besselj{0}\fun{\kappa\rho}
\end{equation}
which form a complete basis in the cylindrical system with invariance along the azimuthal 
and the axial direction. Hence the partial solution to \eqref{eq:cyl:InhomHelmholtz} can
be expanded to the following series
%
\begin{equation}
\left\{
\begin{array}{c}
\hat{I}_n\fun{\rho}\\
B_n\fun{\rho}
\end{array}
\right\}
=
\left\{
\begin{array}{c}
C_{n,0}\\
D_{n,0}
\end{array}
\right\}
+
\sum\limits_{i=1}^{\infty} 
\left\{
\begin{array}{c}
C_{n,i}\\
D_{n,i}
\end{array}
\right\}
\besselj{0}\fun{\kappa_i\rho}
\label{eq:cyl:HMexp}
\end{equation}
with 
%
\begin{equation}
\besselj{0}\fun{\kappa_i\rho_a}=0, \; i=0,1,\ldots\infty.
\end{equation}
The expansion coefficients for the magnetic polarization are calculated by the projections 
onto the basis functions, which for the cylindrical system yields
%
\begin{equation}
C_{n,i} =
\left\{
\begin{array}{ll}
\frac{2}{\rho_a^2}
\int\limits_{0}^{\rho_a} 
\rho \hat{I}_n\fun{\rho} d\rho, 
&$ for $ i=0\\
\frac{2}{E_i^2}
\int\limits_{0}^{\rho_a} 
\rho \besselj{0}\fun{\kappa_i\rho}
\hat{I}_n\fun{\rho,t} d\rho
&$ for $ i=1,2,\ldots\\
\end{array} 
\right.
\end{equation}
where $E_i = \rho_a^2\besselj{1}^2\fun{\kappa_i\rho_a}$.

The corresponding coefficients for the magnetic flux density $D_{n,i}$ are related with 
$C_{n,i}$ via the same relation used in the planar case \eqref{eq:cart:DniCoef}.

Applying the boundary condition at the interface of the rod we obtain the value of the 
$A_n$ coefficient
%
\begin{equation}
\mathcal{K}_0\delta_{n,\pm 1}  = \permb^{-1}
\left[
A_n\besseli{0}\fun{k_n\rho_a} - C_{n,0}
\right].
\end{equation}

The complete solution of the inhomogeneous cylindrical problem for the $n$-th harmonic
can thus be written formally as follows:
%
\begin{eqnarray}
B_n^{(l+1)}\fun{\rho} &=
\permb \mathcal{K}_0\frac{\besseli{0}\fun{k_0\rho}}{\besseli{0}\fun{k_0\rho_a}}\delta_{n,\pm1} 
+C_{n,0}^{(l)} 
\frac{\besseli{0}\fun{k_n\rho}}{\besseli{0}\fun{k_n\rho_a}} 
\nonumber\\
&+ 
\sum\limits_{i=1}^{\infty} 
\frac{\kappa_n^2}{\kappa_i^2 + k_n^2}
C_{n,i}^{(l)}\besselj{0}\fun{\kappa_i\rho}.
\end{eqnarray}
Determining the expansion coefficients $C_{n,i}$ for the unknown magnetization will yield
the magnetic field distribution inside the solenoid.

\subsection{Pulsed excitation}

Passing on the Laplace plane, as we did for the cartesian case, and following the same modal
decomposition as above, we derive the field solution inside the rod for a given magnetic 
polarization
%
\begin{eqnarray}
B\fun{\rho,s} &=
\left[\permb \mathcal{K}_0\fun{s} + C_{n}\fun{s} \right]
\frac{\besseli{0}\fun{\sqrt{sT}\hat\rho}}{\besseli{0}\fun{\sqrt{sT}}}
\nonumber\\ 
&+
\sum\limits_{i=1}^{\infty}
\frac{1}{1+ s\tau_i}
C_{n}\fun{s}\besselj{0}\fun{\kappa_i\rho}
\end{eqnarray}
where this time the characteristic time of the system is given by $T := \permb\cond \rho_a^2$
and the normalized radius by $\hat\rho := \rho/\rho_a$. Decomposing again the coefficient
expressions in terms of rational functions of their poles in the $s$-plane via the GPOF method, 
we obtain
%
\begin{eqnarray}
B\fun{\rho,s} &= \left[
\permb\mathcal{K}\fun{s} + 
\sum\limits_{j=1}^{N_p} \frac{b_{0j} }{s-p_{0j}} \right]
\frac{\besseli{0}\fun{\sqrt{sT}\hat\rho}}{\besseli{0}\fun{\sqrt{sT}}} 
\nonumber\\
&+
\sum\limits_{i=1}^{\infty}
\frac{1}{1+ s\tau_i}
\besselj{0}\fun{\kappa_i\rho}
\sum\limits_{j=1}^{N_p}
\frac{b_{ij}}{s-p_{ij}}.
\label{eq:cyl:RespLD}
\end{eqnarray}

To obtain the time domain solution we work as in the cartesian problem, i.e. we separate the
current and the polarization contributions. 

Let us consider first the inverse Laplace transform of the modified Bessel function ratio 
appearing in \eqref{eq:cyl:RespLD}. In the lemma given in the appendix it is shown that
\begin{equation}
\mathcal{L}^{-1}\left[
\frac{\besseli{n}\fun{\sqrt{sT}\hat\rho}}{\besseli{n}\fun{\sqrt{sT}}} 
\right]
=
\frac{2}{T}
\sum\limits_{k=1}^{\infty}
x_k
\frac{\besseli{n}\fun{x_k\hat\rho}}{\besseli{n}'\fun{x_k}} 
e^{x_k^2 t/T}
\label{eq:cyl:BesselRatInvLap}
\end{equation}
where $x_k$ are the roots of the modified Bessel function $\besseli{0}$. 

The above inverse Laplace transform is easily generalized to include a rational function as follows
\begin{eqnarray}
&\mathcal{L}^{-1}\left[
\frac{1}{s-a}
\frac{\besseli{0}\fun{\sqrt{sT}\hat\rho}}{\besseli{0}\fun{\sqrt{sT}}} 
\right]
\nonumber\\
&\hspace{3ex}
=
\frac{2}{T}
\sum\limits_{k=1}^{\infty}
x_k
\frac{\besseli{0}\fun{x_k\hat\rho}}{\besseli{0}'\fun{x_k}} 
e^{at}\ast e^{x_k^2 t/T}
\end{eqnarray}
which taking into account the relation
\begin{equation}
e^{at}\ast e^{x_k^2 t/T} = \frac{e^{bt}-e^{at}}{b-a}
\end{equation}
becomes
\begin{equation}
\mathcal{L}^{-1}\left[\ldots\right]
=
2
\sum\limits_{k=1}^{\infty}
x_k
\frac{\besseli{0}\fun{x_k\hat\rho}}{\besseli{0}'\fun{x_k}} 
\frac{e^{x_k^2 t/T}-e^{at}}{x_k^2-aT}
\label{eq:cyl:ExpExcit}
\end{equation}
Thus the inverse Laplace transform of the polarization contribution for the $(l+1)$th 
iteration yields
%
\begin{eqnarray}
B_p^{(l+1)}\fun{\rho,t} &= 
\frac{2}{T}
\sum\limits_{j=1}^{N_p} b_{0j}^{(l)}
\sum\limits_{k=1}^{\infty}
x_k
\frac{\besseli{0}\fun{x_k\hat\rho}}{\besseli{0}'\fun{x_k}} 
\frac{e^{x_k^2 t/T}-e^{p_{0j}t}}{x_k^2-p_{0j}T}
\nonumber\\
&+
\sum_{i=1}^{\infty}
\besselj{0}\fun{\kappa_i\rho}
\sum\limits_{j=1}^{N_p}
b_{ij}^{(l)} 
\frac{e^{p_{ij} t} - e^{-t/\tau_i}}{1+p_{ij}\tau_i}
\end{eqnarray}
It is perhaps interesting to note that the zeros of Bessel functions are involved both in the 
spatial ($\kappa_i \rho_a$) and temporal ($x_k$) expansions.

The response to the step current excitation is a special case of \eqref{eq:cyl:ExpExcit} with 
a=0:
\begin{equation}
B_s\fun{\rho,t} =
2\permb\mathcal{K}_0
\sum\limits_{k=1}^{\infty}
\frac{\besseli{0}\fun{x_k\hat\rho}}{\besseli{0}'\fun{x_k}} 
\frac{e^{x_k^2 t/T}-1}{x_k}
\end{equation}
and the response to an arbitrary input waveform will be obtain again by the the Duhamel's 
integral \eqref{eq:cart:Duhamel}.

\section{Results}

The modal solution has been compared with numerical results obtained using a two-dimensional 
numerical code based on the FIT method. The numerical solution is obtained by direct 
integration in the time domain using a two-step backward differentiation implicit Euler scheme 
\cite{clemens_transmag03}.

\begin{figure}[h]
\centering
\includegraphics[width=\figwidth]{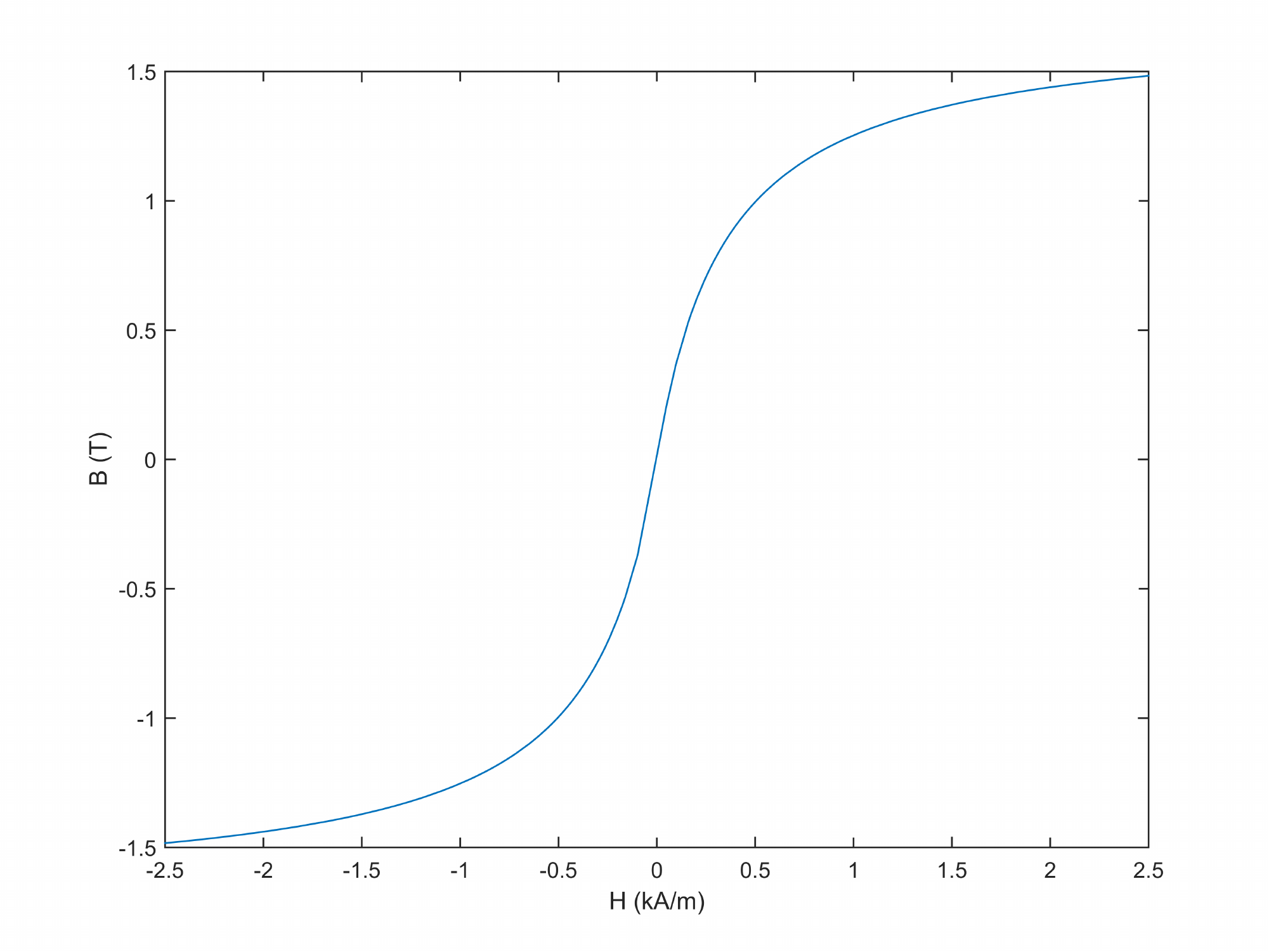}
\caption{Approximated B-H curve of the 1010 steel.}
\label{fig:BHcurve}
\end{figure}

We consider first the configuration of \figref{fig:cart:conf}. The plate thickness is 5~mm and
the material is taken to be 1010 carbon steel with conductivity equal to 6.993~MS/m. When 
working with non-linear material laws it is more convenient to use approximate parametric 
models that directly fit the experimental data. The Fr\"ohlich-Kennelly relation offers such a 
simple yet adequate approximation of the measured non-linear curve, when hysteresis effects 
are neglected. The Fr\"ohlich-Kennelly approximation reads as
%
\begin{equation}
B = \frac{H}{\alpha + \beta |H|}
\end{equation} 
with $\alpha$ and $\beta$ free parameters chosen to best fit the experimental data. The 
estimated values obtained by the fitting procedure for the given steel grade are 
$\alpha=206.42$ and $\beta=0.59148$. The parametric approximation of the material curve is 
shown in \figref{fig:BHcurve}.
%

The non-linear problem has been solved using the two excitations considered above, namely a 
sinusoidal current at $50$~Hz and a step function. The density of the current layer is taken 
equal to 1.5~kA/m for both cases, which lies far beyond the linear region as one can see from
\figref{fig:BHcurve}.  

\begin{figure}[h]
\centering
\includegraphics[width=\figwidth]{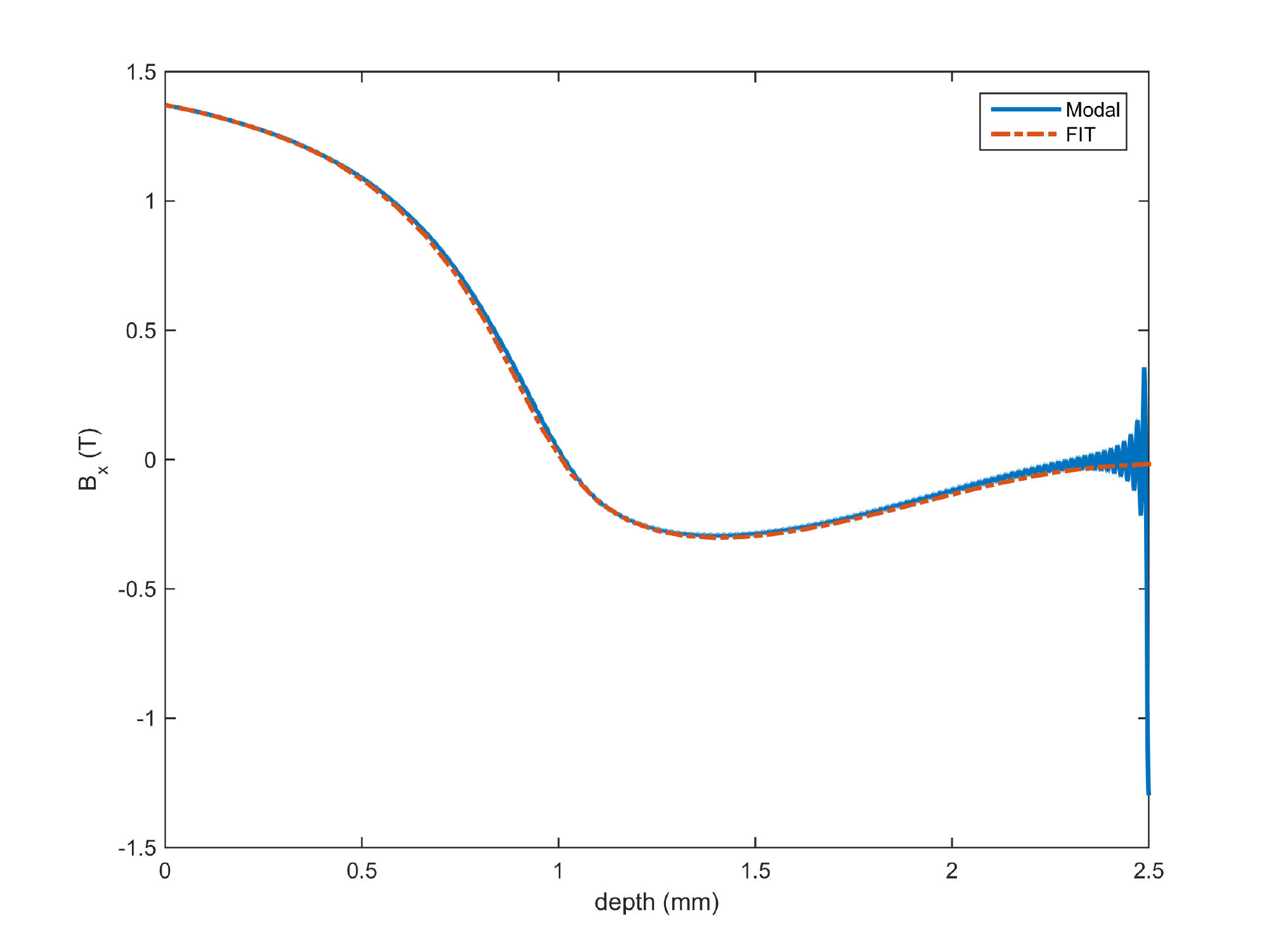}
\\\small{(a)}\\
\includegraphics[width=\figwidth]{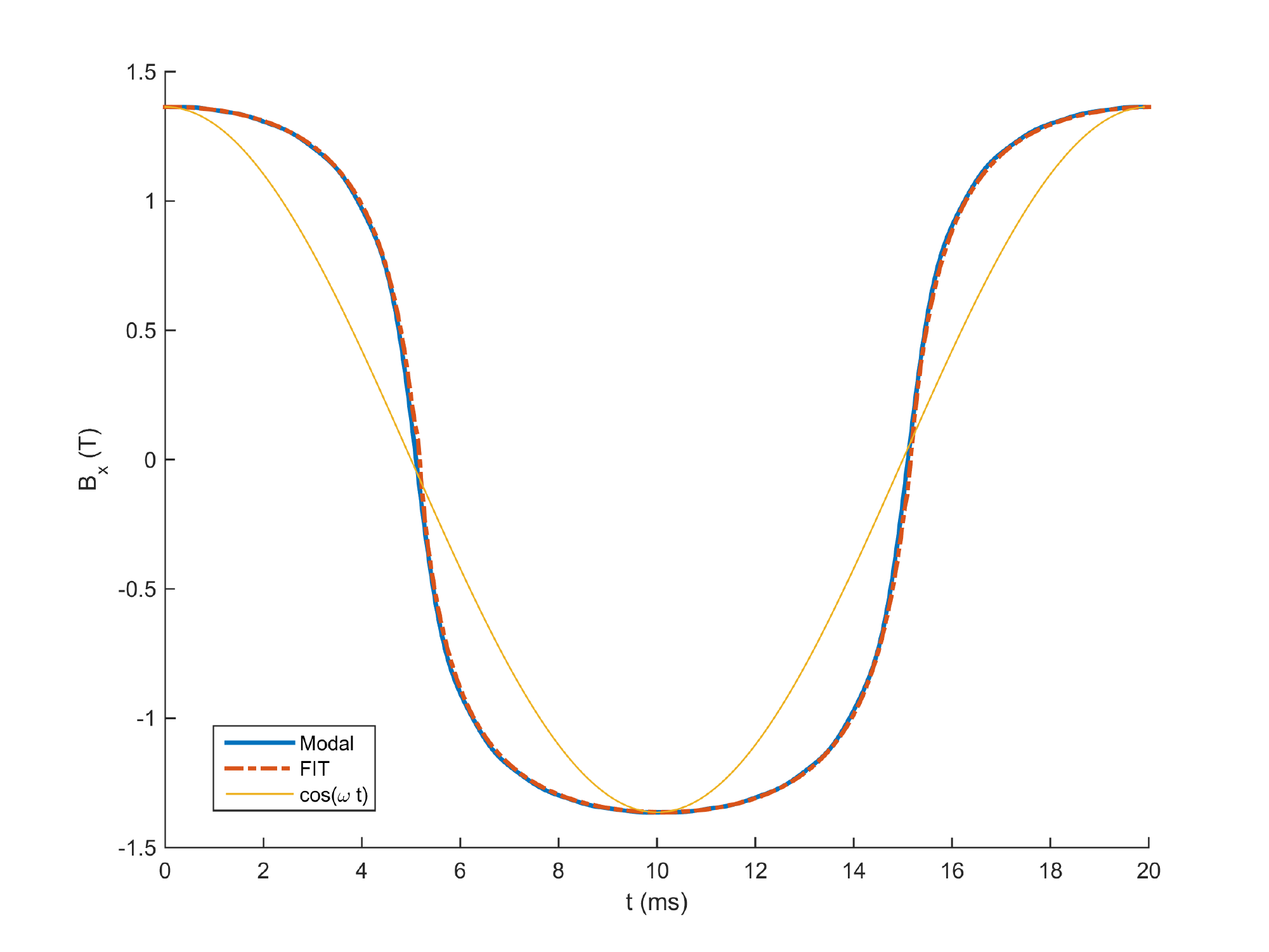}
\\\small{(b)}
\caption{Analytical vs. numerical solution for the magnetic flux density in a ferromagnetic 
plate with harmonic excitation: (a) field profile as a function of depth, (b) time signal at 
25~$\mu$m depth. A harmonic signal of equal amplitude is also drawn to illustrate the
distortion due to the non-linearity.}
\label{fig:BxPlateHrm}
\end{figure}

\Figref{fig:BxPlateHrm}a shows the field variation as a function of the depth at the peak of 
the input. The modal solution is compared with the numerical one obtained with the FIT solver
\cite{clemens_transmag03}. The temporal variation of the magnetic field at a fixed point 
25~$\mu$m underneath the plate surface is depicted in \figref{fig:BxPlateHrm}b. To get a 
picture of the harmonic distortion due to the non-linearity of the material, a harmonic signal 
of the same amplitude is superposed to the field response.
%

\begin{figure}[h]
\centering
\includegraphics[width=\figwidth]{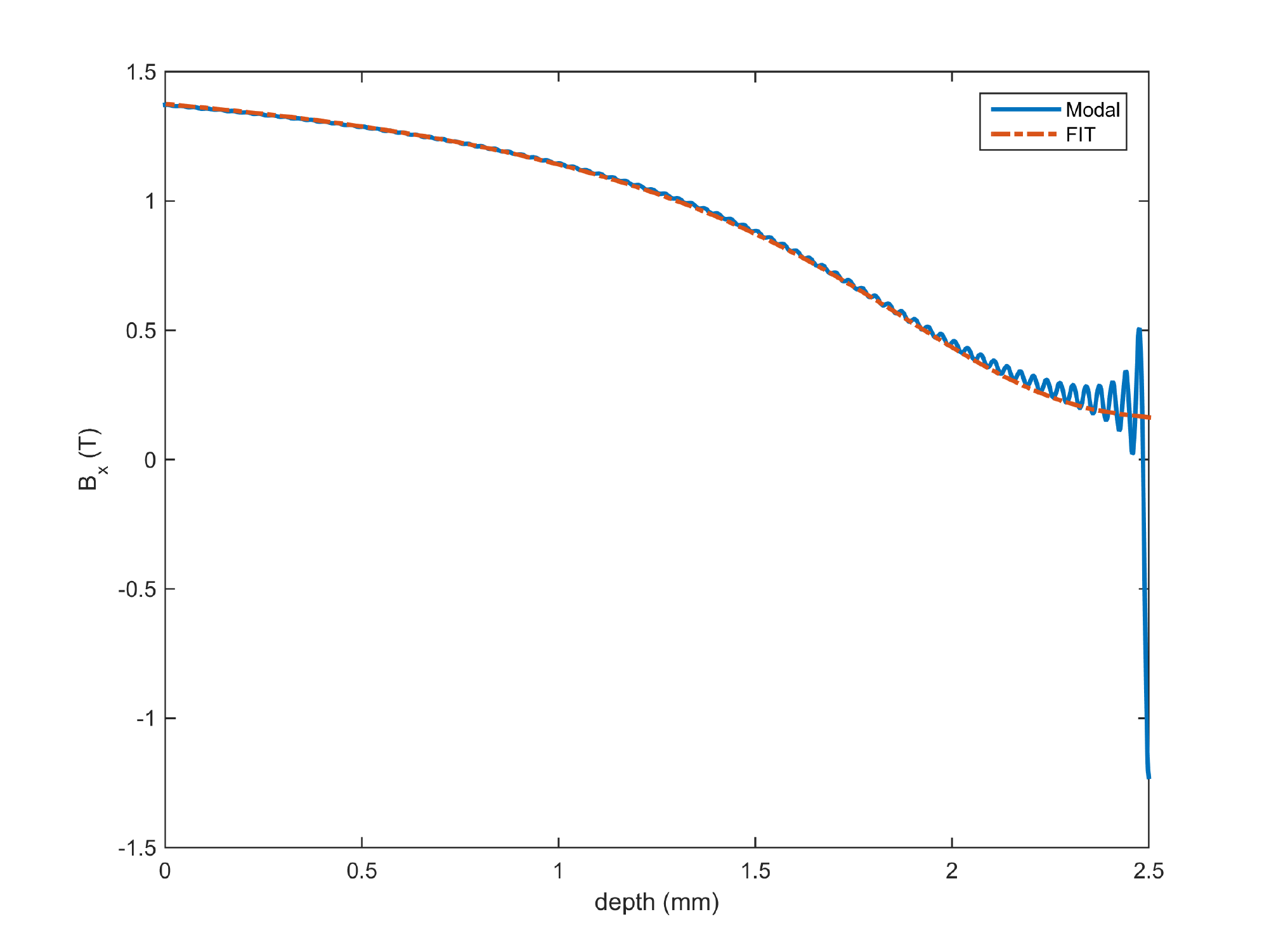}
\\\small{(a)}\\
\includegraphics[width=\figwidth]{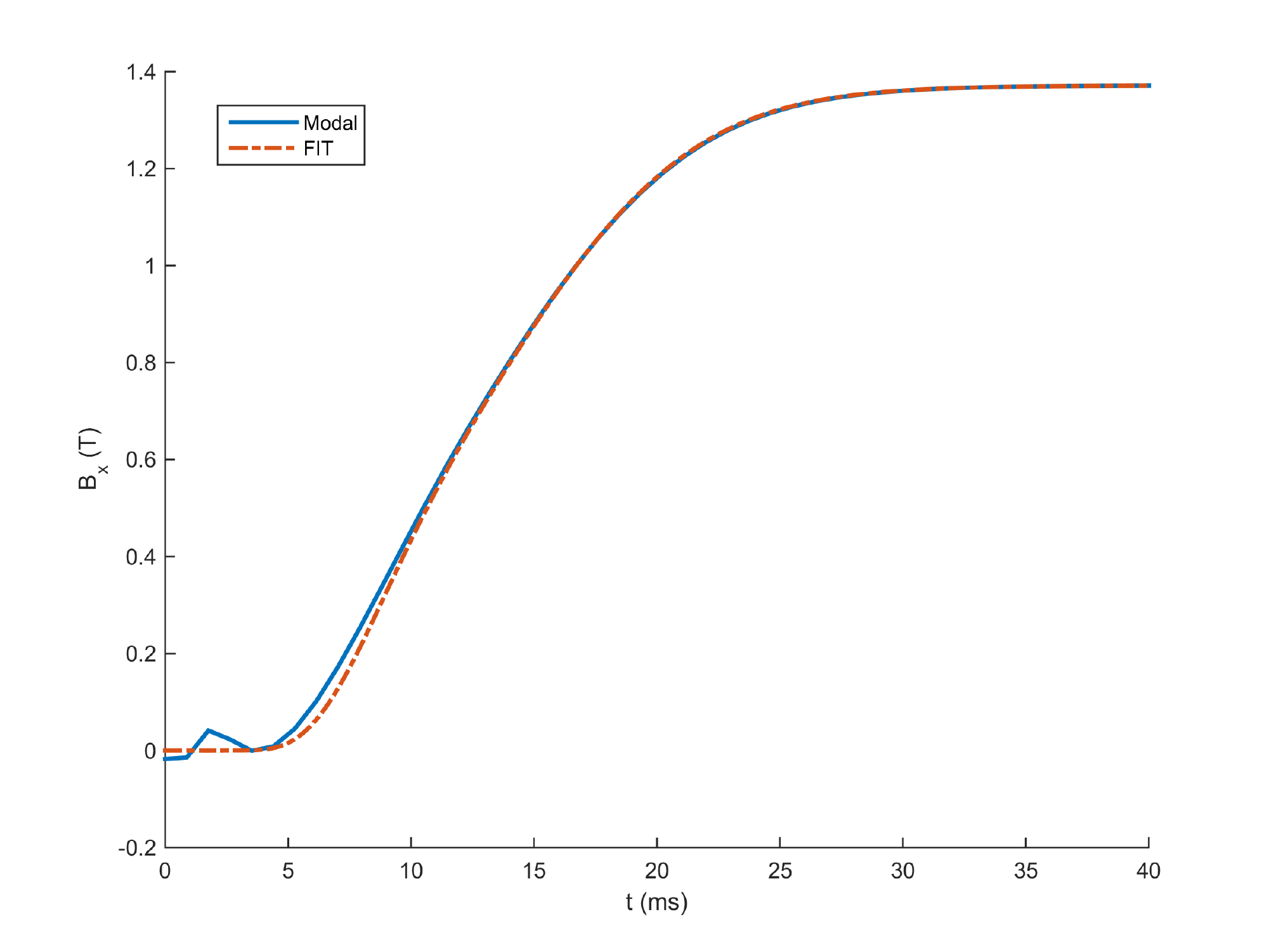}
\\\small{(b)}
\caption{Analytical vs. numerical solution for the magnetic flux density in a ferromagnetic 
plate excited by a step function: (a) field profile as a function of depth, (b) time signal 
at 25~$\mu$m depth.}
\label{fig:BxPlatePls}
\end{figure}

The transient field response to a step excitation is shown in \figref{fig:BxPlatePls}a,b. The 
field profile is given 10~ms after the switching of the current. The field signal is given 
at the same location as in \figref{fig:BxPlateHrm}. 
%

The number of modes taken into account in both calculations (with harmonic and step excitation) 
was 150. The frequency spectrum considered for treatment of the harmonic excitation was 
truncated to the 13th harmonic. As far as the step-function response is concerned, 5 poles 
were proved sufficient to obtain the demonstrated results precision. The computational time 
for the solution of the two problems was 4.2~s and 31~s respectively, which has to be compared 
with the 73~s and 31~s that takes the numerical solution of the same problems using the FIT 
transient solver.

\begin{figure}[h]
\centering
\includegraphics[width=\figwidth]{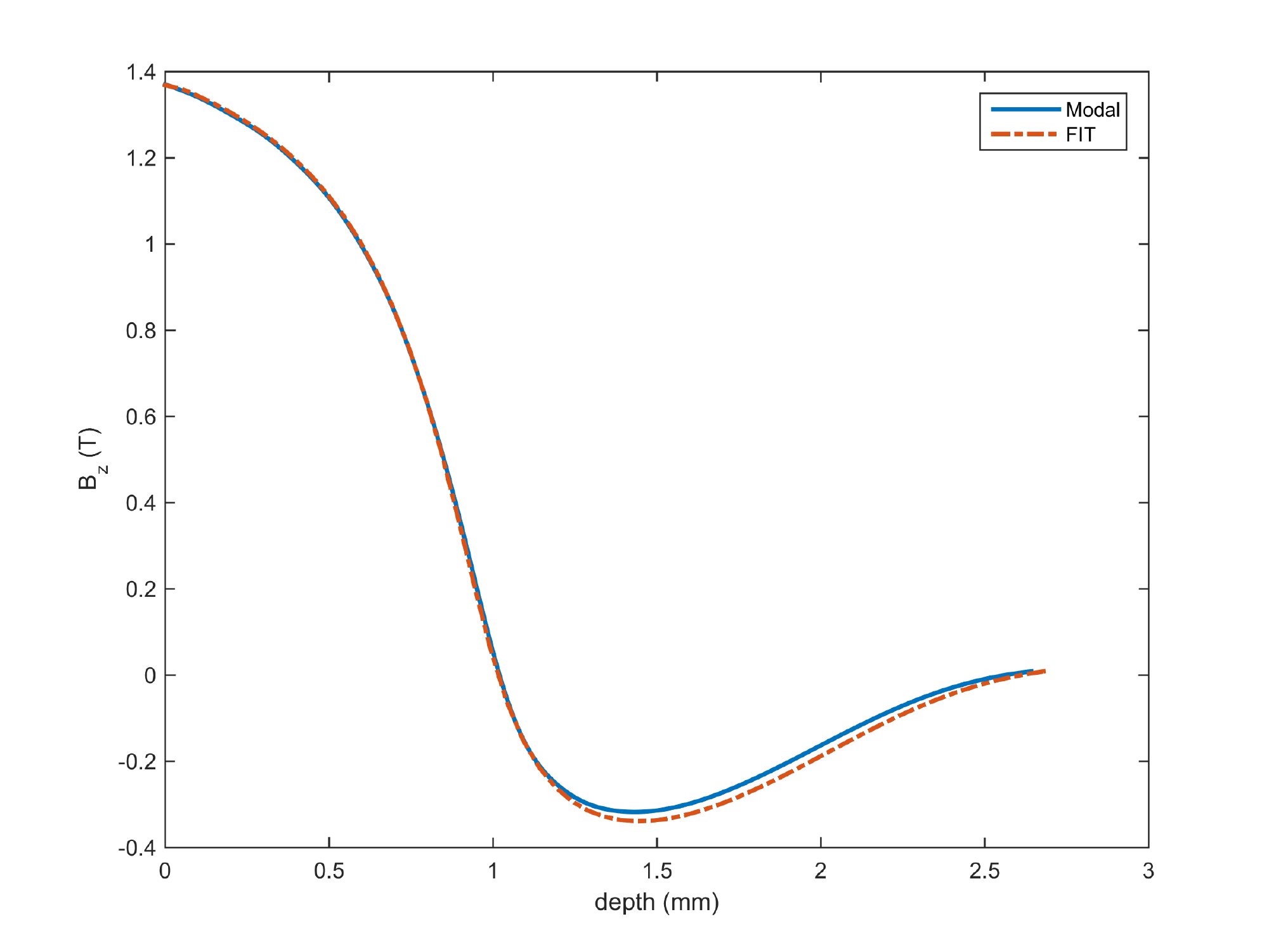}
\\\small{(a)}\\
\includegraphics[width=\figwidth]{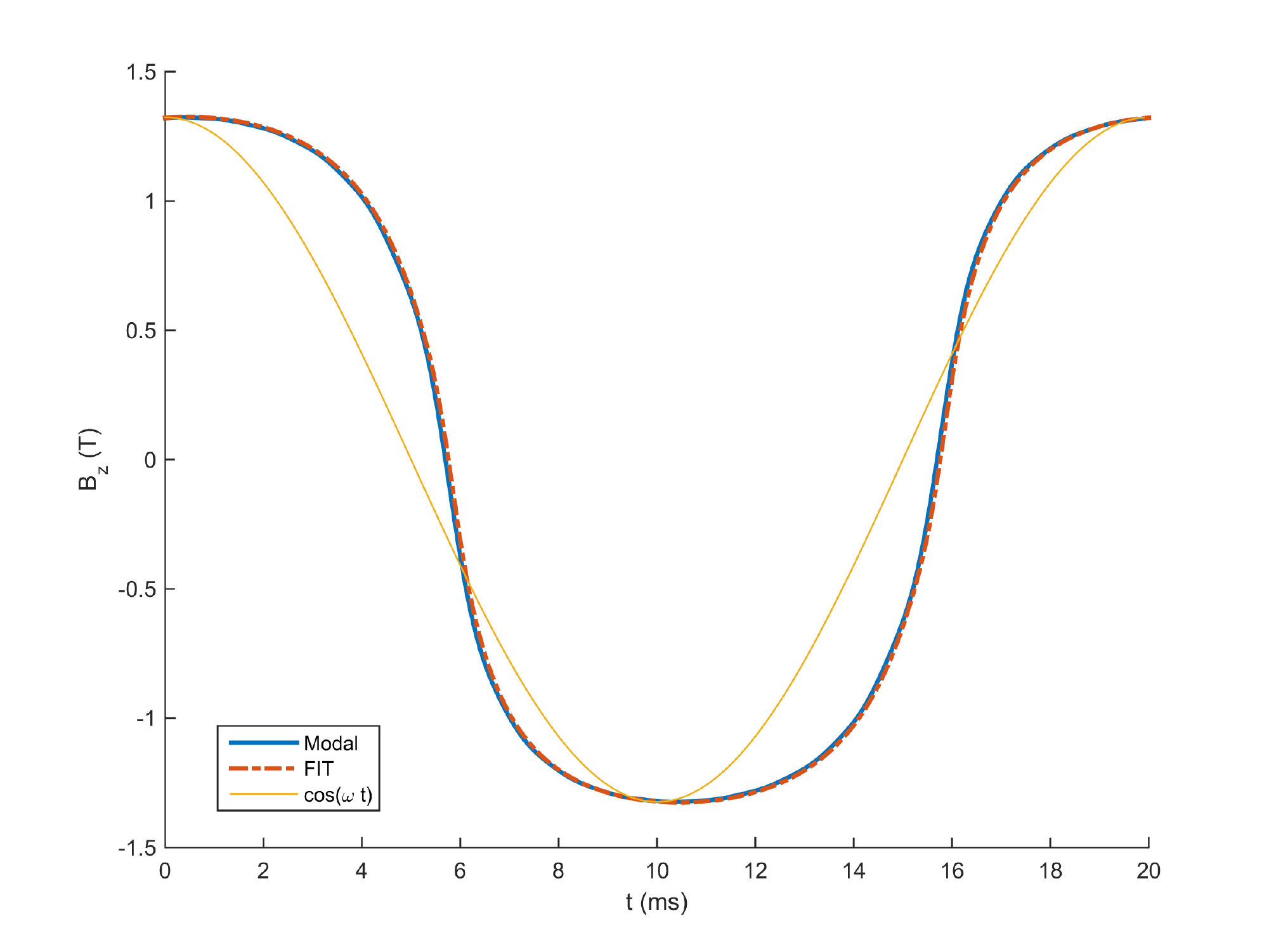}
\\\small{(b)}
\caption{Analytical vs. numerical solution for the magnetic flux density in a ferromagnetic 
rod with harmonic excitation. (a) Field profile as a function of depth, (b) field temporal 
variation.}
\label{fig:BzCylHrm}
\end{figure}

\begin{figure}[h]
\centering
\includegraphics[width=\figwidth]{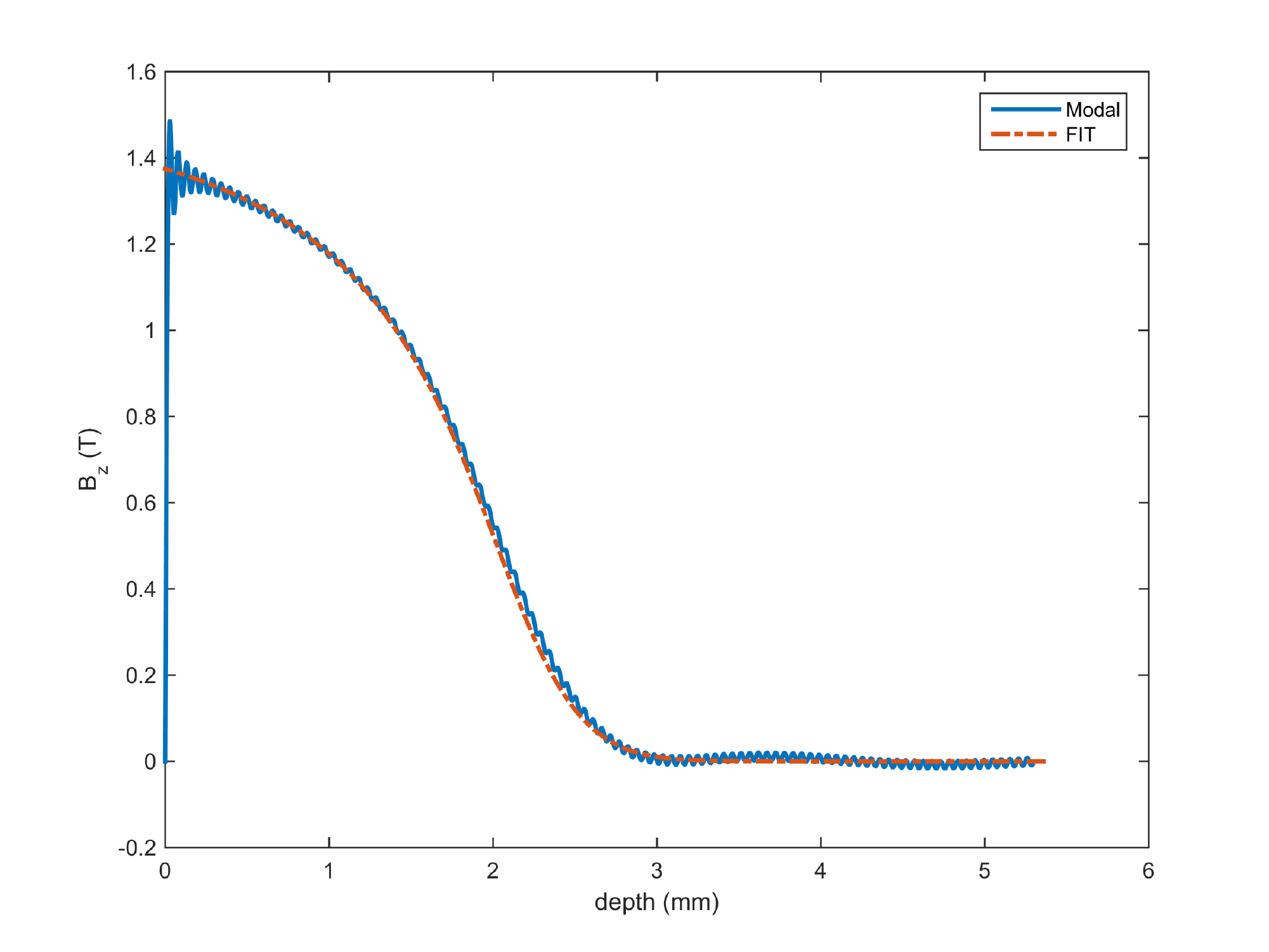}
\\\small{(a)}\\
\includegraphics[width=\figwidth]{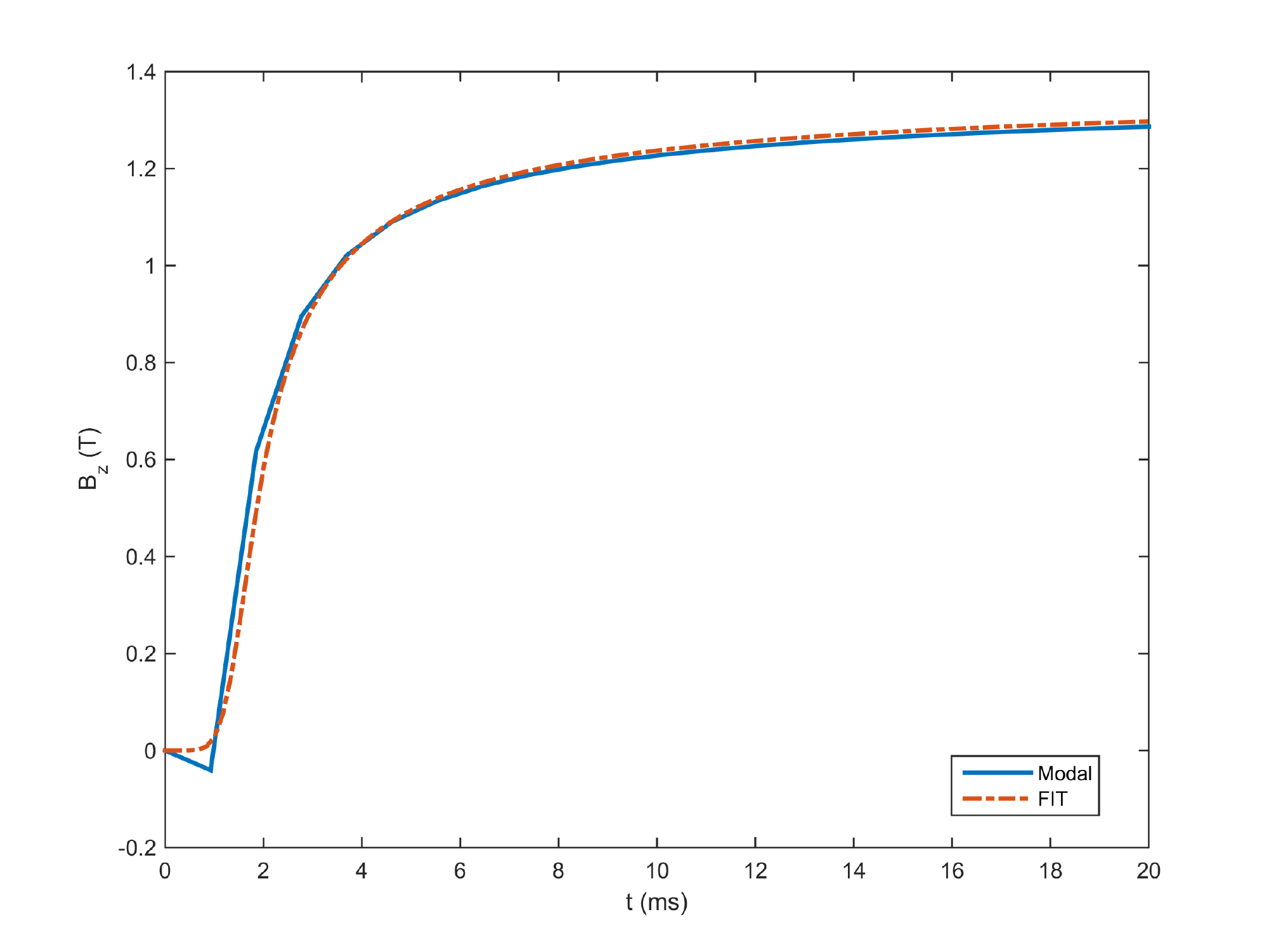}
\\\small{(b)}
\caption{Analytical vs. numerical solution for the magnetic flux density in a ferromagnetic 
rod with harmonic excitation. (a) Field profile as a function of depth, (b) field temporal 
variation.}
\label{fig:BzCylPls}
\end{figure}

The second configuration we consider is the one of the solenoid with the iron core depicted in 
\figref{fig:cyl:conf}. The core diameter is 15.875~mm and consists of the same material as the 
plate (1010 steel). \Figref{fig:BzCylHrm} give the field profile and the time signal for a
harmonic excitation at 50~Hz, whereas \figref{fig:BzCylPls} depicts the corresponding solution 
for step excitation. The field signal in both cases is given at a depth equal to one third of
the rod diameter. The field snapshots are obtained at the peak of the excitation and at 20~ms 
respectively.
%
%

The number of modes considered for the semi-analytical solver was 150 for the harmonic 
excitation and 300 for the step excitation. The number of harmonics taken into account for 
the cosinusoidal excitation was 20, and 12 poles were used for the GPOF decomposition of the
transient problem. The computational times for the modal solution of the two excitations were 
21~s and 70~s using the modal solver, which are clearly lower than the 1205~s and 228~s of the 
numerical solution. 

A general remark that can be made after the comparison of the above computational times is 
that the numerical solution is slower for all the examined cases except for the infinite plate 
with the step excitation, where the two solvers demonstrate the same performances.

In all cases a very satisfactory agreement between the modal and the numerical solution is 
obtained apart from a number of artificial osculations due to the Gibbs effect.

\section{Summary}

The Picard-Banach iterative scheme has been combined with a modal approach for the solution of
the non-linear induction problem in ferromagnetic specimens. For harmonic excitations, the 
enriched spectrum of the higher harmonics is treated by means of Fourier series, whereas the 
transient response to a general pulsed excitation is obtained by applying the Laplace transform. 
The main complication when working in the Laplace domain is the calculation of the inverse 
transforms, which for arbitrary signals has to be done numerically. In this work, an alternative 
approach based on the GPOF algorithm for the signal decomposition in exponential functions is 
used, allowing the calculation of the inverse transform via the superposition of closed-form 
transients.

The present analysis is a first step towards the semi-analytical treatment of non-linear 
problems, and outlines the method that will be employed for the treatment of more realistic 
two and three-dimensional problems. The interest in applying modal solutions rather than 
generic numerical tools lies on the evasion of domain meshing and the reduced computational 
times. 

\section{Acknowledgement}

This work is supported by the CIVAMONT project, aiming at developing scientific collaborations
around the NDT simulation platform CIVA developed at CEA LIST.

\appendices

\section{Asymptotic behaviour for short time}

To study the asymptotic behaviour of the solution for short time, we shall consider the field
response to an exponential pulse (second term of the bracketed expression in 
\eqref{eq:cart:TotalSolPolesDecomp}). The unit step response (first term in the bracket) can 
be considered as a special case of the former with zero pole $p=0$. Thus, let us return to the 
inverse Laplace transform of \eqref{eq:cart:C0InvLaplaceTheta} and let us express it as a 
convolution integral
%
\begin{eqnarray}
s\fun{t}
&:=
\mathcal{L}^{-1}
\left[
\frac{1}{s-p}
\frac{\cosh\fun{x\sqrt{s}}}{\cosh\fun{l \sqrt{s}}}
\right]
\nonumber\\
&=-
\frac{1}{l}
e^{pt} \ast
\frac{\partial}{\partial x}
\thetafun{1}\fun{\frac{x}{2 l}\left|\frac{\iota\pi t}{l^2}\right.}
\end{eqnarray}
where it was made use of \eqref{eq:cart:C0InvLaplaceTheta}. The last relation can also be
written
%
\begin{equation}
s\fun{t}
=-
\frac{1}{2 l^2}
e^{pt} \ast
\frac{\partial}{\partial v}
\left[\thetafun{1}\fun{v\left|\tau\right.}\right]_{v=x/2l,\tau=\iota\pi t/l^2}.
\label{eq:append:convtheta}
\end{equation}

We calculate the derivative of the elliptic theta function making use of its definition 
\eqref{eq:cart:ElThetaFun}
%
\begin{eqnarray}
&\frac{\partial}{\partial v}\thetafun{1}\fun{v\left|\tau\right.} =
(-\iota\tau)^{-1/2}
\sum\limits_{n=-\infty}^{\infty}
(-1)^n e^{-\iota\pi(v-1/2+n)^2/\tau}
\nonumber\\
&=
(-\iota)^{1/2} 2\pi\tau^{-3/2}
\sum\limits_{n=-\infty}^{\infty}
(-1)^n 
(v-1/2+n)
\nonumber\\
&\hspace{3ex}\times
e^{-\iota\pi(v-1/2+n)^2/\tau}
\end{eqnarray}
and we substitute in \eqref{eq:append:convtheta}, which becomes after some simplifications
%
\begin{eqnarray}
s\fun{t}
=
\frac{l}{2\sqrt{\pi}}
&\sum\limits_{n=-\infty}^{\infty}
(-1)^n 
(\frac{x}{l}-1+2n)
e^{pt} \ast t^{-3/2}
\nonumber\\
&\times
e^{-(x/l-1+2n)^2 l^2/4t}.
\label{eq:append:sconvseries}
\end{eqnarray}

Now let us calculate the convolution integral
%
\begin{eqnarray}
I_n\fun{t|a}
&=&
\int\limits_{0}^{t}
e^{p(t-\tau)} \tau^{-3/2}
e^{-a/\tau} d\tau 
\nonumber\\
&=&
e^{pt}
\int\limits_{0}^{t}
e^{-p\tau}
\left[
\frac{d}{d\tau}
\int\limits_{0}^{\tau}
u^{-3/2}e^{-a/u} du
\right]
d\tau.
\end{eqnarray}
With the change of variable $\xi=\sqrt{a/u}$ the interior integral becomes
\begin{equation}
\int\limits_{0}^{\tau}
u^{-3/2}e^{-a/u} du = 
\sqrt{\frac{\pi}{a}} \erfc\fun{\sqrt{\frac{a}{\tau}}}
\end{equation}
whence
\begin{equation}
I_n\fun{t|a}
=
\sqrt{\frac{\pi}{a}}
e^{pt}
\int\limits_{0}^{t}
e^{-p\tau}
\frac{d}{d\tau}
\erfc\fun{\sqrt{\frac{a}{\tau}}}
d\tau
\end{equation}
and applying integration by parts
\begin{eqnarray}
I_n\fun{t|a}
&=
\sqrt{\frac{\pi}{a}}
e^{pt}
\left[
e^{-p\tau}
\erfc\fun{\sqrt{\frac{a}{\tau}}}
\right]_{0}^{t}
\nonumber\\
&+
\sqrt{\frac{\pi}{a}}
pe^{pt}
\int\limits_{0}^{t}
e^{-p\tau}
\erfc\fun{\sqrt{\frac{a}{\tau}}}
d\tau.
\label{eq:append:InInt}
\end{eqnarray}
The integral in the right hand side exists in closed form \cite[7.4.37]{abramowitz}
\begin{eqnarray}
&\int
e^{-p\tau}
\erf\fun{\sqrt{\frac{a}{\tau}}}
d\tau = 
-\frac{1}{p}
\left\{
e^{-p\tau}
\erf\fun{\sqrt{\frac{a}{\tau}}}
\right.
\nonumber\\
&+
\left.
\frac{1}{2}
e^{-p\tau-a/\tau}
\left[
w\fun{\sqrt{-p\tau}+\iota\sqrt{\frac{a}{\tau}}} 
\right.\right.
\nonumber\\
&\left.\left.
\hspace{13ex}+ 
w\fun{-\sqrt{-p\tau}+\iota\sqrt{\frac{a}{\tau}}}
\right]
\right\} + 
\mathrm{const}
\end{eqnarray}
where $w\fun{z} = e^{-z^2}\erfc\fun{-\iota z}$. Hence
\begin{eqnarray}
&\int
e^{-p\tau}
\erfc\fun{\sqrt{\frac{a}{\tau}}}
d\tau =-
\frac{e^{-p\tau}}{p}
\erfc\fun{\sqrt{\frac{a}{\tau}}}
\nonumber\\
&+
\frac{1}{2p}
\left[
e^{2\sqrt{p a}}\erfc\fun{\sqrt{p\tau}+\sqrt{\frac{a}{\tau}}} 
\right.
\nonumber\\
&\left.
\hspace{5ex}+
e^{-2\sqrt{p a}}\erfc\fun{-\sqrt{p\tau}+\sqrt{\frac{a}{\tau}}}
\right]
\end{eqnarray}
where the integration constant is implied. Substituting in \eqref{eq:append:InInt} and taking 
into account $\lim_{z\to\infty} \erfc\fun{z}=0$ we obtain
\begin{eqnarray}
I_n\fun{t|a}
&=
\sqrt{\frac{\pi}{a}}
\frac{e^{pt}}{2}
\left[
e^{2\sqrt{p a}}\erfc\fun{\sqrt{pt}+\sqrt{\frac{a}{t}}} 
\right.
\nonumber\\
&\left.
\hspace{8ex}
+
e^{-2\sqrt{p a}}\erfc\fun{-\sqrt{pt}+\sqrt{\frac{a}{t}}}
\right].
\label{eq:append:InFinal}
\end{eqnarray}

\Eqref{eq:append:sconvseries} can thus be written as 
%
\begin{eqnarray}
s\fun{t}
=
\frac{l}{2\sqrt{\pi}}
&
\sum\limits_{n=-\infty}^{\infty}
(-1)^n 
(\frac{x}{l}-1+2n) 
\nonumber\\
&
\times
I_n\left[t\left|\left(\frac{x-l}{2}+nl\right)^2\right.\right]
\label{eq:append:sfinal}
\end{eqnarray}
i.e. we have expressed the inverse Laplace transform as an infinite series of complementary 
error functions.

Particular interest presents the short time limit, i.e. $t\to 0$. The series of 
\eqref{eq:append:sfinal} is a rapidly converging series, which means that for small t (or 
equivalently for large complementary error function arguments) we can consider only the zero 
term ($n=0$):
\begin{equation}
s\fun{t}
\approx 
\frac{1}{2\sqrt{\pi}}(x-l) 
I_n\left[t\left|\frac{(x-l)^2}{4}\right.\right]
\end{equation}
which substituting \eqref{eq:append:InFinal} and setting $x\rightarrow \sqrt{T}\hat{z}$, 
$l\rightarrow \sqrt{T}/2$\footnote{Only the positive sign of the quantity $x-l$ has physical 
meaning.} yields
\begin{eqnarray}
&s\fun{t}
\approx 
\frac{e^{pt}}{2}
\left[
e^{\sqrt{pT}(1/2-\hat{z})}\erfc\fun{\sqrt{\frac{T}{4t}}(1/2-\hat{z})+\sqrt{pt}} 
\right.
\nonumber\\	
&\hspace{3ex}+
\left.
e^{-\sqrt{pT}(1/2-\hat{z})}\erfc\fun{\sqrt{\frac{T}{4t}}(1/2-\hat{z})-\sqrt{pt}}
\right].
\nonumber\\
\label{eq:append:sFinal}
\end{eqnarray}

Now let us consider the field response to an exponential signal $e^{pt}$ in a half-space. The 
response in the Laplace domain is given by the relation
\begin{equation}
B_e\fun{z,s} \sim 
\frac{e^{-\sqrt{s\cond\permb}(d/2-z)}}{s-p} = 
\frac{e^{-\sqrt{s T}(1/2-\hat{z})}}{s-p}
\end{equation}
which recalling the transform \cite[(10), p. 246]{erdelyitabitrans} is inverted into the time 
domain as follows
\begin{eqnarray}
&B_e\fun{z,t}
\sim
\frac{e^{p t}}{2}
\left[
e^{\sqrt{pT}(1/2-\hat{z})} 
\erfc\fun{\sqrt{\frac{T}{4t}}(1/2-\hat{z}) + \sqrt{pt}}
\right.
\nonumber\\
&\left.
\hspace{3ex}+
e^{-\sqrt{p T}(1/2-\hat{z})} 
\erfc\fun{\sqrt{\frac{T}{4t}}(1/2-\hat{z}) - \sqrt{pt}}
\right]
\end{eqnarray}
which is identical with \eqref{eq:append:sFinal}. In other words the plate response for short
time reduces to the one of the half-space. From the point of physical interpretation, this 
means that at sufficiently short time, the eddy-current field has not yet reached the $z=-d/2$ 
boundary, and therefore the plate is equivalent to a half-space.

\section{Derivation of \eqref{eq:cyl:BesselRatInvLap}}

We seek the inverse Laplace transform of the complex function
%
\begin{equation}
F\fun{s} = \frac{\besseli{n}\fun{\sqrt{sT}\hat\rho}}{\besseli{n}\fun{\sqrt{sT}}}.
\end{equation}
This transform is not found in the most standard inverse transform tables. 

To examine the behaviour of $F\fun{s}$ in the $\mathrm{Re}\left[s\right]>0$ half-plane for
$s\rightarrow\infty$, we consider the asymptotic form of the modified Bessel functions
\begin{equation}
\frac{\besseli{n}\fun{\sqrt{sT}\hat\rho}}{\besseli{n}\fun{\sqrt{sT}}} \sim 
\exp\left[\sqrt{sT}(\hat\rho-1)\right]/\sqrt{\hat{\rho}}
\end{equation} 
for $\left|\arg\right|<\pi/2$, which is analytic and vanishes at $s\rightarrow\infty$ for 
every $\hat\rho<1$. Hence $F\fun{s}$ is analytic and $\lim_{z\to\infty} F\fun{s} = 0$, bounded 
in the $\mathrm{Re}\left[s\right]>0$ half-plane except for the zeros of the Bessel function in 
the denominator, and the inverse Laplace transform exists. For $\hat\rho=1$, $F\fun{s}=1$, and 
its inverse transform is the Dirac delta function. 

Let us consider $\hat\rho<1$. Using a basic result of the complex analysis (known as complex 
inversion formula \cite{marsdenbook}) the inverse Laplace transform of the function $F\fun{s}$ 
can be written as
%
\begin{equation}
f\fun{t} = \sum\limits_{k=1}^{\infty} \mathrm{Res}\left[F\fun{s} e^{st}\right]_{s=s_k}
\label{eq:append:CompInvFormula}
\end{equation}
where $\mathrm{Res}\left[F\fun{s} e^{st}\right]$ are the residues of $F\fun{s} e^{st}$. 
Substituting the considered transformed function \eqref{eq:cyl:BesselRatInvLap} into 
\eqref{eq:append:CompInvFormula}, we obtain
\begin{eqnarray}
&\mathcal{L}^{-1}\left[
\frac{\besseli{n}\fun{\sqrt{sT}\hat\rho}}{\besseli{n}\fun{\sqrt{sT}}} 
\right]
=
\sum\limits_{k=1}^{\infty}
\frac{\besseli{n}\fun{\sqrt{s_k T}\hat\rho}}
{\frac{\partial \besseli{n}\fun{\sqrt{s_k T}}}{\partial s}} 
e^{s_k t}
\nonumber\\
&\hspace{3ex}=
\frac{1}{2 \sqrt{T}}
\sum\limits_{k=1}^{\infty}
\sqrt{s_k}
\frac{\besseli{n}\fun{\sqrt{s_k T}\hat\rho}}
{\besseli{n}'\fun{\sqrt{s_k T}}} 
e^{s_k t}
\label{eq:append:invLTransf}
\end{eqnarray}
where $s_k$ is related to the Bessel function zeros $x_k$ via $\sqrt{s_k T} = x_k$. 
\Eqref{eq:append:invLTransf} thus becomes 
\begin{equation}
\mathcal{L}^{-1}\left[\ldots\right]
=
\frac{2}{T}
\sum\limits_{k=1}^{\infty}
x_k
\frac{\besseli{n}\fun{x_k\hat\rho}}{\besseli{n}'\fun{x_k}} 
e^{x_k^2 t/T}
\end{equation}
which is the sought relation.



\end{document}